\documentclass[conference,compsoc]{IEEEtran}

\usepackage{algorithmic}
\usepackage{algorithm}

\makeatletter
\newcommand{\removelatexerror}{\let\@latex@error\@gobble}
\makeatother
\usepackage{booktabs} 
\usepackage{amsmath} 
\usepackage{multirow}
\usepackage{adjustbox}
\usepackage{enumitem}
\usepackage{amsfonts}
\usepackage{url}
\usepackage{cite}

\ifCLASSINFOpdf
\else
\fi

\begin{document}
\title{EMK-KEN: A High-Performance Approach for Assessing Knowledge Value in Citation Network}

\author{
\IEEEauthorblockN{Zehui Qu\textsuperscript{1}\textsuperscript{*}}
\IEEEauthorblockA{School of Software, \\School of Computer and \\
Information Science\\
Southwest University\\
Chongqing, China\\
quzehui@swu.edu.cn\\
Telephone: (86) 18323886531\\
Fax: (86) 83252352}
\and
\IEEEauthorblockN{Chengzhi Liu\textsuperscript{1}\textsuperscript{*}}
\IEEEauthorblockA{School of Software, \\School of Computer and \\
Information Science\\
Southwest University\\
Chongqing, China 400700\\
liuchengzhi@email.swu.edu.cn}
\and
\IEEEauthorblockN{Hanwen Cui\textsuperscript{2}}
\IEEEauthorblockA{School of Software, \\School of Computer and \\
Information Science\\
Southwest University\\
Chongqing, China\\
1737279629@qq.com}
\and
\IEEEauthorblockN{Xianping Yu\textsuperscript{3}}
\IEEEauthorblockA{School of Software, \\School of Computer and \\
Information Science\\
Southwest University\\
Chongqing, China\\
yuxp1018@swu.edu.cn}
}

\maketitle

\begin{abstract}
With the explosive growth of academic literature, effectively evaluating the knowledge value of literature has become quite essential. However, most of the existing methods focus on modeling the entire citation network, which is structurally complex and often suffers from long sequence dependencies when dealing with text embeddings. Thus, they might have low efficiency and poor robustness in different fields. To address these issues, a novel knowledge evaluation method is proposed, called EMK-KEN. The model consists of two modules. Specifically, the first module utilizes MetaFP and Mamba to capture semantic features of node metadata and text embeddings to learn contextual representations of each paper. The second module utilizes KAN to further capture the structural information of citation networks in order to learn the differences in different fields of networks. Extensive experiments based on ten benchmark datasets show that our method outperforms the state-of-the-art competitors in effectiveness and robustness.
\end{abstract}

\begin{figure}[b]
\vskip -6mm
\begin{tabular}{p{44mm}}
\toprule\\
\end{tabular}
\vskip -4.5mm
\noindent
\setlength{\tabcolsep}{1pt}
\begin{tabular}{p{1.5mm}p{79.5mm}}
$1$& first author, * corresponding author.
\end{tabular}
\end{figure}

\IEEEpeerreviewmaketitle

\section{Introduction}
Evaluation of knowledge value is a challenge in academic fields, especially for recently published papers. Currently, the evaluation system can be divided into two categories: paper citation counts and citation network-based evaluation\cite{HOU2019100197}. However, both of them focus on entire citation networks to improve accuracy. This will lead to excessive computational costs. 

Traditional paper citation metrics, such as the H-index\cite{AnIndexToQuantify}, impact factor\cite{garfield1972citation}, and citation frequency, assess academic influence from their respective singular perspectives. For example, the H-index considers the scholarly output and the frequency of citations, while the impact factor primarily measures the influence of the journal. However, these methods predominantly rely on citation frequency, which may overlook a vast number of papers that would make significant contributions.

In the context of evaluation through citation networks, researchers initially turned to statistical models. Yan et al. (2011) \cite{yan2011citation} developed a method to forecast the future frequency of citation of the literature using various regression models, such as linear regression and support vector regression, which effectively captured the popularity characteristics of the literature. Trujillo and Long (2018) \cite{co-citationanalysis} employed a bibliographic coupling analysis methodology to construct and validate citation networks, demonstrating the application of statistical modeling for knowledge evaluation and interdisciplinary research network modeling in scientific literature. Anil et al. (2020) \cite{anil2020effect} introduced a metric to assess the class imbalance in heterogeneous networks and examined its influence on the prediction of collaborators and the classification of the research domain of the author. Although these methods are effective for prediction of numerical data, they cannot be used to process structured information in citation networks of academic papers.

To address these challenges, researchers have turned to neural network architectures. For instance, Qin et al. (2017) \cite{qin2017dual} introduced a recursive neural network (DA-RNN) with a two-stage attention mechanism, which effectively predicts time series data and shows potential for dynamic prediction of the citation network. Zhang and Wu (2020) \cite{zhang2020predicting} proposed the WMR Rank framework, which utilizes seven types of relationships in dynamic heterogeneous networks to predict the future influence of papers, researchers and academic venues. Li, Tang et al. (2022) \cite{li2022predicting} developed a four-layer multilayer perceptron model to predict the clinical citation frequency of biomedical papers, emphasizing the importance of the characteristics of the reference dimension in the prediction. Gao et al. (2024) \cite{3637871} developed SMLP4Rec, an efficient all-MLP architecture for sequential recommendation. By parallelizing MLP4Rec's structure and adding normalization layers, it improves training efficiency and prediction accuracy, offering new perspectives for knowledge evaluation and data mining in neural network-based models. These studies highlight the potential of neural networks to process complex data and predict future trends, reinforcing the feasibility of using neural networks for the modeling of citation networks. However, although models based on traditional neural networks are adept at processing data with nonlinear relationships, they struggle to describe and understand data with network structures.

In order to better understand the structural information of citation networks, graph neural networks (GNNs) have been introduced to describe and analyze the topological information between nodes and edges \cite{Wang2025}. For example, Wu et al. (2023)\cite{10026520} systematically reviewed medical knowledge graphs (MKGs), analyzing their data sources, construction methods, reasoning models, and healthcare applications. They proposed a Reasoning Implementation Pathway (RIP) for MKG classification and application analysis, providing theoretical and practical insights for medical knowledge evaluation and data mining. He et al. (2023) \cite{he2023h2cgl} constructed H2CGL, combining hierarchical and heterogeneous graph models, and employed contrastive learning to enhance the sensitivity of the graph representation to potential citations, thereby effectively predicting the potential impact of a paper. Yang Yu et al. (2023) \cite{yu2021incorporating} proposed a novel joint extraction model that combines BiLSTM and GCN to capture both sequential and structural semantic features of sentences. Zou et al. (2023) \cite{zou2023se} introduced the SE-GSL framework, which provides a universal and effective approach to graph structure learning through structural entropy optimization, thereby enhancing the model's robustness in noisy and heterogeneous structures. Duan et al. (2024) \cite{duan2024structural} proposed a novel Graph Neural Network (GNN) framework based on structural entropy theory, which enhances node classification performance by extracting hierarchical community information through an encoding tree mechanism to optimize graph structure representation. Hong et al. (2024) \cite{hong2024label} proposed the label attention distillation method LAD-GNN, which addresses the embedding issue in graph classification tasks by training a teacher model and a student GNN. 

Despite the advantages of GNNs in the acquisition of structural information from citation networks, traditional GNN models typically require modeling of the entire citation network, which can lead to high computational demands and lengthy training times, making it difficult to process large-scale citation networks\cite{corso2024graph, zhou2020graph}. Moreover, GNN models are highly sensitive to parameters and often lack robustness and generalization capabilities when dealing with literature networks of different domains\cite{gamarnik2023barriers, uddin2024}.

To address this challenge, this work proposes a novel method of knowledge evaluation called the \textbf{Efficient Mamba-KAN Knowledge Evaluator (EMK-KEN)}, which combines the Mamba and KAN architectures. It possesses the advantages of both of the architectures. These advantages include the complexity of the linear data processing time, the ability to perform selective information propagation/forgetting, and the efficient handling of long sequence data that Mamba possesses\cite{SchiffKGDGK24, gu2023mamba, behrouz2024graph, dao2024transformers}, as well as the smaller parameters and scale, fine-grained and diverse nonlinear transformations, and the powerful expressive ability that KAN has\cite{SHUKLA2024117290, liu2024kan, liu2024kan2, bresson2024kagnns, kiamari2024gkan, yang2024kolmogorov}. The approach begins by using Mamba to process the literature metadata and text embeddings separately, then generate feature vectors. After that, the output vectors from the central node are passed to KAN for further learning. Using KAN’s learnable activation function, the characteristic outputs are generated for downstream tasks, such as classification and citation prediction. To verify the accuracy of the model, multiple classification methods are tested on ten benchmark datasets from different academic fields. The experimental results demonstrate that the EMK-KEN method proposed in this paper is significantly superior to traditional GNNs. Additionally, the experiments of parameter sensitivity and multi-index analyses are conducted for deep evaluation. The main contributions of this work are as follows:

\begin{itemize}
\item \textbf{Rapid Processing of Network Structure Data:} Theoretically and empirically, the Kolmogorov-Arnold representation theorem possess faster neural scaling laws than MLPs. The linear time complexity of the Mamba model offers computational advantages over traditional GNN models, enabling the EMK-KEN approach to provide faster real-time predictions and assessments for large-scale citation networks. The method proposed in this paper features a more streamlined structural design. By constructing the network topology using only the literature and its direct references, it reduces complexity and enhances training efficiency and prediction speed.
\item \textbf{Boosting Prediction Accuracy:} Kolmogorov-Arnold representation theorem confirmed that smaller KANs can achieve comparable or better accuracy than larger MLPs in evaluating tasks of knowledge value from citation networks. The 10 benchmark datasets for the citation network from various fields in KAN were used to evaluate various classification methods and compare them with six other models. The experimental results demonstrated the effectiveness and superiority of EMK-KEN.
\item \textbf{Enhancing Generalization Ability:} The selective propagation and forgetting mechanism of Mamba enables the EMK-KEN model to effectively handle citation network features across different domains. It efficiently extracts global relationships within the network and the context of citation relationships while filtering out structurally irrelevant information. This approach aims to improve the model's generalization capability.
\end{itemize}

\section{Model}
\subsection{General introduction}
\begin{figure*}[!t]
\centering 
\includegraphics[width=7.00in]{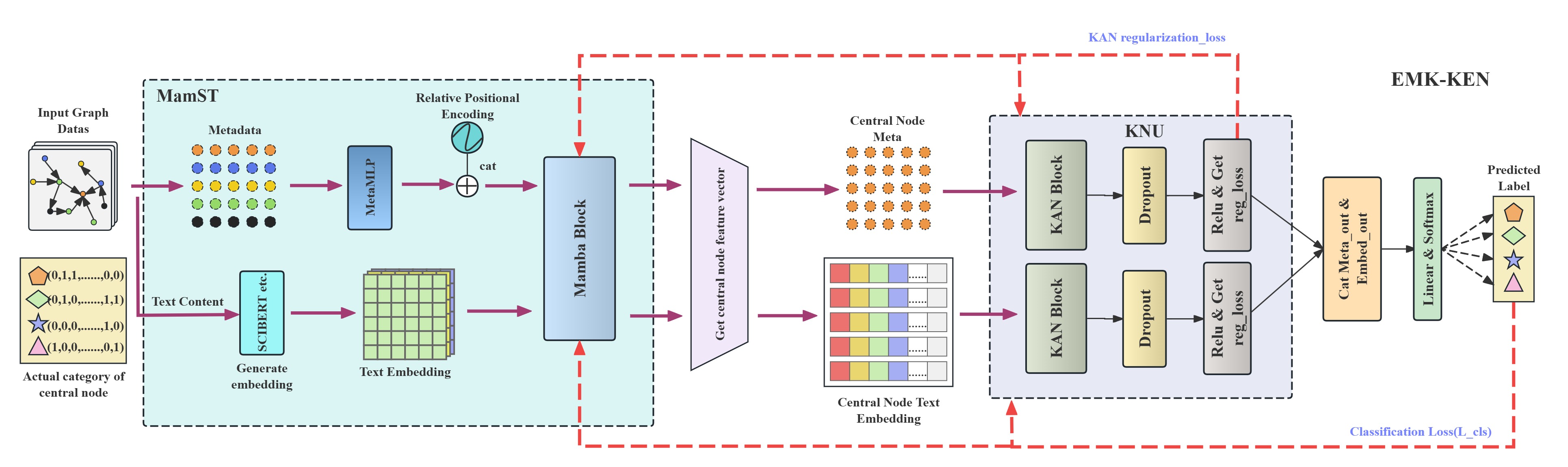}
\caption{EMK-KEN structure}
\label{fig:model1}
\end{figure*}

EMK-KEN is to classify the knowledge level of literature based on the citation network graph structure and metadata. The output is a high-dimensional vector representing the knowledge level of the literature, which can then be utilized as input for downstream tasks. To achieve this, this model is proposed and constructed, which consists of the Mamba Long Order Processor (MambaSeq Transformer, MamST) and the KANflex Neural Unit (KNU).

The workflow of the model, as depicted in Figure \ref{fig:model1}, is outlined as follows. Initially, the model receives graph data as input. Subsequently, it utilizes the MamST network to discern the inherent relationships embedded within the graph-structured data. This process involves handling the feature attributes of each node, which encompass both metadata and long-sequence text embeddings. Following this, the KNU module processes the outputs from the central nodes of the MamST. The KNU is composed of a layer of KAN, a regularization function, and a loss function. Within KNU, metadata and text embeddings are independently processed to produce distinct outputs. In the final stage, the EMK-KEN module integrates these outputs, processes them through a linear layer, and applies the Softmax function to derive the prediction probabilities for each classification category.

The initial step involves the utilization of MamST to handle structured graph data. For each graph, the input features are denoted as $\boldsymbol{X} \in \mathbb{R}^{N \times F}$, where $N$ means the number of nodes, and $F$ represents the quantity of the feature. The citation matrix is expressed as $\boldsymbol{E}$. To comprehensively exploit the information from both metadata and texts, distinct methodologies have been employed for each data type. The subsequent sections delineate the complete processing workflows for metadata and text data, respectively.

\subsection{Metadata Processing}
\subsubsection{Initialization}
\begin{figure}[h]
\centering 
\includegraphics[width=3.30in]{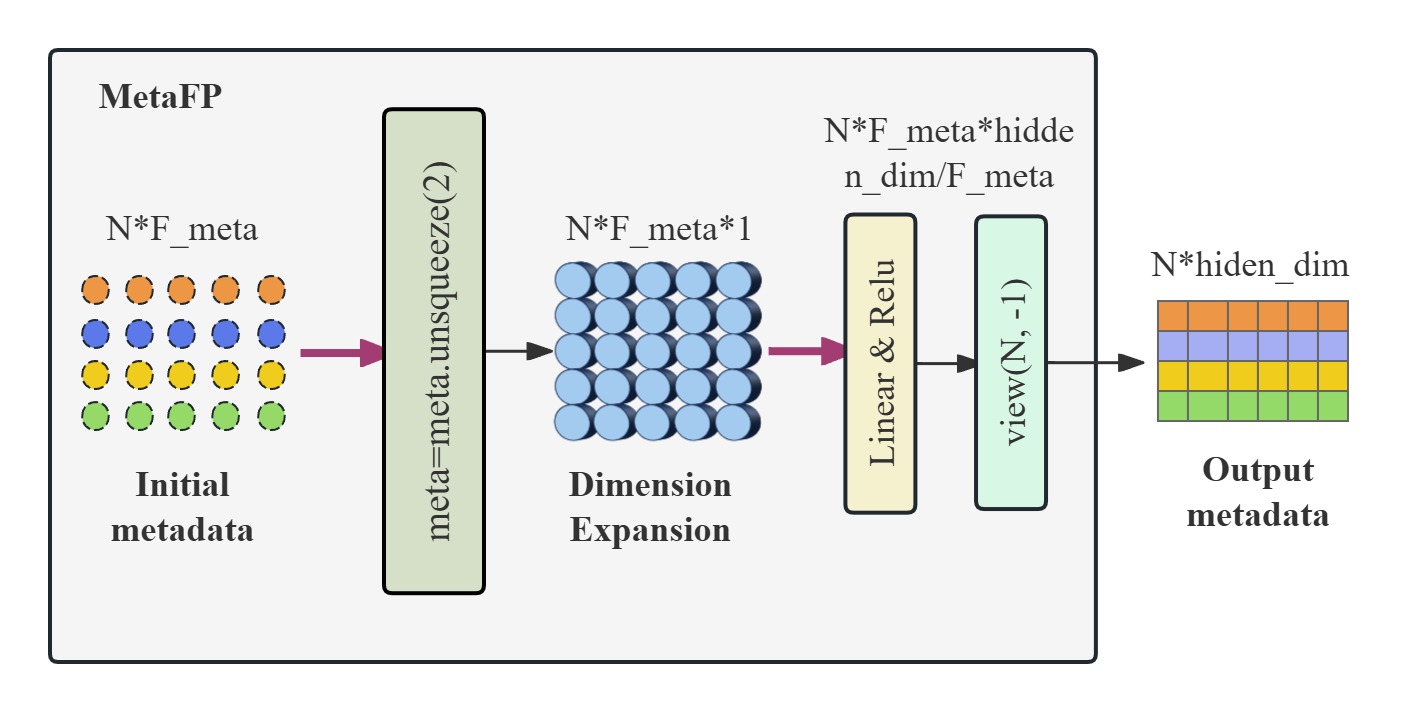}
\caption{MetaFP structure}
\label{fig:MetaFP}
\end{figure}

For each graph, the input for each node is represented as $\boldsymbol{X}=[\boldsymbol{meta}, \boldsymbol{embedding}]$, where $\boldsymbol{meta} \in \mathbb{R}^{N \times F_{\text{meta}}}$ and $\boldsymbol{embedding} \in \mathbb{R}^{N \times seq\_len \times F_{\text{embed}}}$. Among them, seq\_1en is the length of the text embedding sequence. Since the metadata input for nodes often exhibits sparsity, the model initially preprocesses the input data to enhance its learning capability. As shown in Figure \ref{fig:MetaFP}, the metadata is processed through MetaFeature Preprocessor (MetaFP). The specific processing steps are as follows: \\
Firstly, the tensor undergoes a transformation from a two-dimensional to a three-dimensional structure, specifically, its dimensionality changes from $N \times F_{\text{meta}}$ to $N \times F_{\text{meta}} \times 1$. Subsequently, the tensor is processed through a linear layer followed by an activation function (as depicted in Equation \ref{eq:1}), resulting in a new tensor with dimensions $N \times F_{\text{meta}} \times H_{\text{dim}} / F_{\text{meta}}$. Following this, the second and third dimensions of the resulting tensor are combined to produce the refined metadata, represented as $\text{meta}_{\text{fc}} \in \mathbb{R}^{N \times H_{\text{meta}}}$, where $H_{\text{meta}} = \frac{H_{\text{dim}}}{F_{\text{meta}}}$.

\begin{equation}
        \boldsymbol{meta}_{\text{fc}}=\text{ReLU}\left(\boldsymbol{meta} \cdot \boldsymbol{W}_{\text{fc}}+\boldsymbol{b}_{\text{fc}}\right)
	\label{eq:1}
\end{equation}
Here, $\boldsymbol{W}_{\text{fc}} \in \mathbb{R}^{1 \times H_{\text{meta}}}$ represents the weight matrix, and $\boldsymbol{b}_{\text{fc}} \in \mathbb{R}^{H_{\text{meta}}}$ denotes the bias vector. ${H_{\text{dim}}}$ corresponds to the dimension of the hidden layer in MetaFP, while ${H_{\text{meta}}}$ represents the dimension of the output layer.

\subsubsection{Relative Position Encoding and MamST}
To enhance the predictive accuracy of the model, Relative Position Encoding is employed, and the MamST is utilized to reduce computational load, significantly improving computational efficiency. Following initialization, the output tensor is concatenated with the relative position encoding. Building upon this, the information from the position encoding is superimposed onto the original data, which is subsequently processed as input to Mamba. The dimension of the relative position encoding is specified as $\mathbb{R}^{N \times 1}$. As illustrated in Equation \ref{eq:2}, the merged metadata is obtained, with a shape of $\mathbb{R}^{N \times \left(H_{\text{meta}} + 1\right)}$. Through the linear activation layer (Equation \ref{eq:3}), the data are transformed into a shape that conforms to the input requirements of the Mamba layer, specifically $\mathbb{R}^{N \times seq\_len \times \left(H_{\text{meta}} + 1\right)}$. Here, $\text{seq\_len}$ denotes the number of nodes in the citation network graph.
\begin{equation}
        \boldsymbol{meta}_{\text{cat}}=\text{Concat}\left(\boldsymbol{meta}_{\text{fc}}, \boldsymbol{pos\_encoding}\right)
	\label{eq:2}
\end{equation}
\begin{equation}
        \boldsymbol{meta}_{\text{comb}}=\text{ReLU}\left(\boldsymbol{meta}_{\text{cat}}\cdot \boldsymbol{W}_{\text{metafc}}+\boldsymbol{b}_{\text{metafc}}\right)
	\label{eq:3}
\end{equation}
\par
The computational workflow within the Mamba layer involves several key steps: linear transformation, causal convolution, state space model (SSM) processing, and final output calculation. The following section elaborates on these processing steps in detail.

In the initial linear transformation phase, Mamba firstly reshapes the input data from $(N, seq\_len, H_{\text{meta}}+1)$  to $(H_{\text{meta}}+1, N \times seq\_len)$. Subsequently, a weight matrix $\boldsymbol{W}_{\text{in\_proj}}$ is applied to perform a linear transformation, expanding the feature dimension from one dimension to two dimensions. That is, the resulting tensor is then reshaped from $(H_{\text{meta}}+1, N \times seq\_len)$ to $(N, 2(H_{\text{meta}}+1), seq\_len)$. Furthermore, a bias term $\boldsymbol{b}_{\text{in\_proj}}$ is added to the result (as shown in Equation \ref{eq:4}). All this linear transformation step serves the purpose of mapping the input features into a higher-dimensional space (2D) and divide it into two parts, thereby facilitating subsequent convolution and state-space model computations.
\begin{equation}
        \boldsymbol{xz}=\boldsymbol{W}_{\text{in\_proj}} \cdot   
        \begin{pmatrix}
             N\\
             seq\_len\\
             H_{\text{meta}}+1  
       \end{pmatrix}     
      +\boldsymbol{b}_{\text{in\_proj}}
	  \label{eq:4}  
\end{equation}

Subsequently, apply causal convolution. At this stage, the obtained tensor $\boldsymbol{xz}$ is first divided into two parts along the last dimension, resulting in two components $\boldsymbol{x}$ and $\boldsymbol{z}$, each with a shape of $(N, H_{\text{meta}}+1, seq\_len)$ \ref{eq:5}. For the obtained $\boldsymbol{x}$, causal convolution is carried out via the function causalConv1dFn (Equation \ref{eq:6}).
\begin{equation}
    \boldsymbol{x}, \boldsymbol{z} = \text{chunk}(\boldsymbol{xz}, 2, \text{dim}=-1) 
    \label{eq:5}
\end{equation}
\begin{equation}
        \boldsymbol{x}'=\text{causalConv1dFn}\left(\boldsymbol{x}, \boldsymbol{W}_{\text{conv1d}}, \boldsymbol{b}_{\text{conv1d}},  \text{SiLU}\right)
	\label{eq:6}
\end{equation}
Among them, $\boldsymbol{W}_{\text{conv1d}}$ is the weight of the convolution kernel, with a shape of $(2(H_{\text{meta}}+1), 1, W)$, where $W$ is the width of the convolution kernel. $\boldsymbol{b}_{\text{conv1d}}$ is the bias term of the convolutional kernel, with a shape of $(2(H_{\text{meta}}+1))$. $\text{SiLU}$ is the activation function.

The data is then fed into the State Space Model (SSM) for processing. Initially, the input feature $\boldsymbol{x}'$ is linearly projected via the matrix $\boldsymbol{W}_{\text{x\_proj}}$ to obtain the time step ($\Delta t$), state matrix $\boldsymbol{B}$ and output matrix $\boldsymbol{C}$ (Equation \ref{eq:7}). Based on the obtained state matrix $\boldsymbol{B}$ and the state transition matrix $\boldsymbol{A}$, the state $S_{\text{state}}$ is updated. This step represents the core operation of the SSM and simulates the evolution of the state over time. The mathematical equation for this process is shown in Equation \ref{eq:8}. Here, matrix $\boldsymbol{A}$ describes the relationship between the system states. Specifically, it is a linear transformation matrix used to predict the state at the next time step. Subsequently, using the updated state $S_{\text{state}}$ and the input $\boldsymbol{x}'$, the output y is obtained by applying the state matrix $\boldsymbol{C}$ and an additional trainable parameter D (Equation \ref{eq:9}). The parameter D represents the skip connection parameter, which is used to control the weight during the state update process. Its shape is ${(\text{inner}})$. The skip connection is a mechanism that allows the network to directly pass input to subsequent layers, thereby mitigating the vanishing gradient problem and enhancing the model's expressive power\cite{SchiffKGDGK24}. Finally, the output $\boldsymbol{y}$ is combined with the activated $\boldsymbol{z}$ (obtained by applying an activation function), and this result is further integrated with the input feature $\boldsymbol{z}$ to update the output $\boldsymbol{y}$ of the state space model.
\begin{equation}
        \Delta t, \boldsymbol{B}, \boldsymbol{C}=\text{split}(\boldsymbol{W}_{\text{x\_proj}} \cdot \boldsymbol{x}')
	\label{eq:7}
\end{equation}
\begin{equation}
        S_{\text{state\_next}}=\boldsymbol{A} \cdot S_{\text{state}}+\Delta t \cdot (\boldsymbol{B} \cdot \boldsymbol{x}')
	\label{eq:8}
\end{equation}
\begin{equation}
     \boldsymbol{y} = (S_{\text{state\_next}} \cdot \boldsymbol{C} + \boldsymbol{D} \cdot \boldsymbol{x}') \odot \text{SiLU}(\boldsymbol{z})
    \label{eq:9}
\end{equation}
Herein, matrices $\boldsymbol{A}$, $\boldsymbol{B}$, $\boldsymbol{C}$, and the state vector $S_{\text{state}}$ have the following respective shapes: (d\_inner, d\_state), (batch\_size, d\_state, seq\_len), (batch\_size, d\_state, seq\_len), and (batch\_size, d\_inner, d\_state). The parameter d\_inner is obtained by multiplying the input feature dimension by the expansion factor (default is 2) and rounding up, representing the expanded internal feature dimension. d\_state on the other hand, is a crucial parameter that determines the shapes of both the state matrix and state vector. In the subsequent section, the impact of this parameter on model training will be analyzed.

Output calculation is performed via the output layer. Then it passes through a Dropout layer to mitigate overfitting. Afterward, the shape of the data is adjusted to $\mathbb{R}^{N \times\left(H_{\text{meta}}+1\right)}$.

\subsubsection{KANflex Neural Unit}
Then, EMK-KEN will extract the central node metadata from the total output of Mamba. This metadata  is then fed into the KNU for further processing. In KNU, the input data is denoted as $\boldsymbol{X} \in \mathbb{R}^{N \times H}$. Here $N$ represents the size of the batch and $H$ is the dimension of the features. The tensor $\boldsymbol{X}$ applies the weight matrix $\boldsymbol{W}_{\text{base}}$, similar to a traditional fully connected network (Equation \ref{eq:10}). Next, B-spline basis function B(x) is computed by grid calculations, and then multiplied by spline weights $\boldsymbol{W}_{\text{spline}}$ to perform B-spline interpolation (Equation \ref{eq:11}). At the same time, a weight scaler (spline scaler) is used (Equation \ref{eq:12}) to dynamically adjust the weight values of spline interpolation, thereby enhancing the model's generalization performance. The final output is a combination of the results of the linear transformation and spline interpolation (Equation \ref{eq:13}). Using the spline interpolation algorithm, the proposed approach exhibits superior characterization ability for simultaneous linear and nonlinear transformations. Notably, the tensor produced in this phase has a dimensionality of $N \times \frac{KNU_{\text{Hdim}}}{2}$. This tensor is then processed through the Dropout layer prior to output generation, effectively mitigating potential overfitting issues.

\begin{equation}
        \boldsymbol{baseOutput} =\boldsymbol{W}_{\text{base}} \cdot \boldsymbol{x}+\boldsymbol{b}_{\text{base}}
	\label{eq:10}
\end{equation}
\begin{equation}
        \boldsymbol{splineOutput}=\boldsymbol{W}_{\text{spline}} \cdot B(x)
	\label{eq:11}
\end{equation}
\begin{equation}
        \boldsymbol{W}_{\text{scaled\_spline}}=\boldsymbol{W}_{\text{spline}} \times splineScaler
	\label{eq:12}
\end{equation}
\begin{equation}
        \boldsymbol{output} = \text{Dropout}(\boldsymbol{baseOutput} + \boldsymbol{splineOutput})
	\label{eq:13}
\end{equation}

\subsection{Text Data Processing }
For textual data processing, the model implements the following pipeline: Initially, pre-trained language models (such as SciBERT \cite{beltagy2019scibert}, CharBERT \cite{ma2020charbert}, and MatSciBERT \cite{gupta2022matscibert}) are utilized to process the textual content, generating corresponding text embeddings \cite{Zhang2021}. These embeddings are subsequently processed by the MamST. The processing mechanism resembles that of metadata, with a notable distinction: text embeddings bypass the MetaFP and relative position encoding modules. Following MamST processing, the output is directed through a dropout layer. Subsequently, the central node representation is extracted by pooling the average sentences of the text embeddings, producing a tensor $\mathbb{R}^{N \times 1 \times F_{\text{embed}}}$. This tensor is then reshaped to $\mathbb{R}^{N \times F_{\text{embed}}}$ before being fed into the KNU (Knowledge Neural Unit). Within the KNU, text embeddings undergo the same processing procedures as metadata embeddings.

\subsection{Final output}
After obtaining the central node metadata and text embeddings, EMK-KEN concatenates them to form a feature vector with dimensions of $N \times 2 \times KNU_{\text{outputdim}}$. This vector takes as the input of the final linear layer, transforming it into a result of dimension $N \times N_{\text{Class}}$. EMK-KEN applies the Softmax function to the vector, yielding the classification probability. The algorithm 1 presents the complete workflow of the model.

\begin{figure}[!t]
\label{algo1}
\renewcommand{\algorithmicrequire}{\textbf{Input:}}  
\renewcommand{\algorithmicensure}{\textbf{Output:}}  
\removelatexerror
\begin{algorithm}[H]
\caption{EMK-KEN Algorithm}
\begin{algorithmic}[1]
\REQUIRE Graph datasets $D$, labels $Y$, hyper parameter $\lambda$.
\ENSURE Optimized model parameters, predicted graph labels $\hat{Y}$.
\STATE Randomly initializes model parameters.
\FOR{each training epoch $t$} 
    \FOR{each batch $b$} 
        \STATE Preliminary training of metadata using MetaFP, followed by merging metadata  and positional encoding:
        \STATE \hspace{1em} $\boldsymbol{H}_{\text{meta}} = \boldsymbol{W}_{\text{fc}} F_{\text{meta}} + \boldsymbol{b}_{\text{fc}}$
        \STATE \hspace{1em} $\boldsymbol{H}_{\text{init}} = \boldsymbol{H}_{\text{meta}} + \text{PosEnc}(X_{\text{graph}})$
        \STATE Using Mamba block to cope with metadata and text embedding separately:
        \STATE \hspace{1em} $\boldsymbol{H}_{\text{Mamba}} = \text{Dropout}(\sigma(\boldsymbol{W}_{\text{Mb}} \boldsymbol{H}_{\text{init}} + \boldsymbol{b}_{\text{Mb}}))$
        \STATE The central node mask is used to identify which node serves as the central node. Based on the central node mask ($Cmask$), the features of the central node ($\boldsymbol{H}_{\text{cMamba}}$) are extracted and processed by the KNU, respectively.
        \STATE \hspace{1em} $\boldsymbol{H}_{\text{cMamba}} = \text{GetCenter}(\boldsymbol{H}_{\text{Mamba}}, Cmask)$
        \STATE \hspace{1em} $H_{\text{KAN}} = \sigma(\boldsymbol{W}_{\text{KAN}} \boldsymbol{H}_{\text{cMamba}} + \boldsymbol{b}_{\text{KAN}})$
        \STATE Concatenate outputs of both branches:
        \STATE \hspace{1em} $\boldsymbol{H} = [\boldsymbol{H}_{\text{MetaKAN}}; \boldsymbol{H}_{\text{EmbedKAN}}]$
        \STATE Linear layer for classification:
        \STATE \hspace{1em} $\hat{Y} = \text{Softmax}(\boldsymbol{W}_{\text{out}} \boldsymbol{H} + \boldsymbol{b}_{\text{out}})$
        \STATE Compute loss and update parameters:
        \STATE \hspace{1em} $L = L_{\text{CLS}}(\hat{Y}, Y) + \lambda \cdot KAN.\text{reg\_loss}$
    \ENDFOR
\ENDFOR
\end{algorithmic}
\end{algorithm}
\end{figure}

\section{Dataset and experiment}
\begin{table*}[!t]
  \centering
  \caption{\normalfont Dataset statistics}
    \resizebox{\textwidth}{!}{ 
    \begin{tabular}{ccccccccccc}
    \toprule
         & ACMHypertextECHT & Oxytocin & ACMIS & SNAP-HEP-TH & ALTEGRADChallenge2021 & OGBN-ArXiv & Citation Networks-V12 & DBLP-V13 & AMN  & Biomedicine \\
    \midrule
    Node & 728  & 1467 & 16148 & 27770 & 100336 & 169343 & 4894081  & 5354309 & 60646 & 31546 \\
    Edges & 5326 & 86535 & 128027 & 352807 & 2633688 & 1166243 & 45564149 & 48277950 & 264505 & 211883 \\
    Features of Each Node & 771  & 770  & 768  & 768  & 768  & 128  & 769  & 768  & 772  & 772 \\
    avg\_degree\_mean & 2.028565999 & 3.209401105 & 1.68724248 & 5.739597091 & 2.27999313 & 3.209401105 & 2.121065134 & 3.291421 & 2.753928 & 2.98286728 \\
    density\_mean & 0.339621773 & 0.309399259 & 0.33165093 & 0.281348555 & 0.267120259 & 0.309399259 & 0.174349121 & 0.160286 & 0.136204 & 0.11100418 \\
    clustering\_coefficient\_mean & 0.311047304 & 0.447885524 & 0.17575515 & 0.589518431 & 0.280954154 & 0.447885524 & 0.197611676 & 0.418762 & 0.158879 & 0.15973088 \\
    merged\_avg\_degree & 6.874861573 & 27.50422833 & 6.25811098 & 25.40921858 & 15.7684171 & 9.618699808 & 8.74428356 & 8.07193843 & 5.932137 & 5.89177895 \\
    merged\_density & 3.81E-03 & 7.27E-03 & 1.19E-04 & 4.58E-04 & 5.69E-05 & 4.48E-05 & 2.26E-06 & 1.14E-05 & 6.72E-05 & 7.08E-05 \\
    merged\_clustering\_coefficient & 0.15810383 & 0.212251555 & 0.14831729 & 0.312019496 & 0.16797002 & 0.501645854 & 0.064397687 & 0.140762852 & 0.470022 & 0.48720974 \\
    Classes & 3    & 7    & 3    & 2    & 2    & 4    & 2    & 2    & 2    & 2 \\
    \bottomrule
    \end{tabular}%
  \label{tab:dataset-statistics}%
  }
\end{table*}%
In this study, a comprehensive analysis of multiple citation network datasets was conducted to explore the mutual citation relationships and academic influence among academic literature. Specifically, experiments were carried out using eight developed datasets: (1) SNAP-HEP-TH\cite{snapnets}, (2) ACMHypertextECHT\cite{ACMHypertextECHT}, (3) ALTEGRADChallenge2021\cite{ALTEGRADChallenge2021}, (4) Citation Networks-V12\cite{Citation-Networks-V12, sinha2015overview}, (5) DBLP-V13\cite{DBLP-V13, Tang:08KDD, Tang:10TKDD, Tang:11ML, Tang:12TKDE, Tang:07ICDM}, (6) ACMIS\cite{DVN_27695_2014}, (7) OGBN-ArXiv\cite{OGBN-ArXiv, wang2020microsoft, mikolov2013distributed}, and (8) Oxytocin\cite{leng_2022_6615221}. These datasets provided rich citation network structures and node feature information.

Based on these, citation networks were constructed for each paper and its references, and features—including metadata and text embeddings were added to each node. Subsequently, three classification methods were employed for testing. For datasets (1) to (6), the reference network was classified based on the in-degree of the central node, as well as the number of nodes and edges in the reference network, before being fed into the model for prediction \cite{IdentifyingNode, sinha2015overview, Tang:08KDD, Tang:10TKDD, Tang:11ML, Tang:12TKDE, Tang:07ICDM}. For datasets (7) to (8), the provided classification criteria were directly utilized for the classification experiments \cite{OGBN-ArXiv, wang2020microsoft, mikolov2013distributed, leng_2022_6615221}. Additionally, the Knowledge Quantization Index (KQI)\cite{xu2011quantifying, fanelli2019theory}, based on graph structure entropy, was adopted as the classification criterion.

Finally, a total of 60,646 articles from three subfields of computer science were collected to construct the (9) AMN citation network dataset. These papers are all from the fields of Artificial Intelligence, Machine Learning, and Natural Language Processing, and their publication dates fall between January 1, 1970, and December 31, 2003. Similarly, 31,546 articles from three subfields of biomedical science were also collected, forming the (10) Biomedicine dataset. These three subfields are Bioinformatics, Biotechnology, and Physiology, and their publication dates align with those of the AMN dataset. The basic information of each dataset is shown in Table \ref{tab:dataset-statistics}.

The concept of the Knowledge Quantification Index (KQI) is as follows: the KQI gauges the degree to which each piece of literature impacts the overall structure of the knowledge network by quantifying the variation in the structural entropy of the citation network\cite{ye2014approximate, coutrot2022entropy, xu2022methodology, yang2023minimum, Jin2024}. Its theoretical foundation is derived from the concept of graph structural entropy\cite{zhang2024human, yang2024incremental}. This entropy value not only considers the probability distribution of nodes but also incorporates the hierarchical community structure of the nodes in the network. In the context of a given multi-inheritance knowledge tree $T$, the Equation for KQI is expressed as follows:
\begin{equation}
        \kappa_{\alpha}^{T}=-\sum_{1 \leq i \leq d_{\alpha}^{i n}} \frac{V_{\alpha}^{T}}{d_{\alpha}^{in} W} \log _{2} \frac{V_{\alpha}^{T}}{d_{\alpha}^{in} V_{\alpha_{i}^{-}}^{T}}
	\label{eq:14}
\end{equation}
Where:
\begin{equation}
        V_{\alpha}^{T}=d_{\alpha}^{out}+\sum_{1 \leq i \leq d_{\alpha}^{out}} \frac{V_{\alpha_{i}^{+}}^{T}}{d_{\alpha_{i}^{+}}^{in}}
	\label{eq:15}
\end{equation}
In the Equation \ref{eq:14}, \ref{eq:15}, $\kappa_{\alpha}^{T}$ signifies the KQI value of node $\alpha$ within the knowledge tree $T$. $\alpha_{i}^{-}$ stands for the i-th parent node of node $\alpha$. $\alpha_{i}^{+}$ denotes the i-th child node of node $\alpha$. $V_{\alpha}^{T}$ represents the aggregate of all $V_{\alpha}$ in the various fragments of the knowledge tree $T$. $V_{\alpha_{i}^{-}}^{T}$ indicates the volume of the i-th parent node of node $\alpha$ in the knowledge tree $T$. $W$ serves as the total volume of the entire knowledge tree, which is employed to normalize the KQI value. $d_{\alpha}^{i n}$ represents the in-degree of node $\alpha$, namely, the number of parent nodes that node $\alpha$ has. $d_{\alpha_{i}^{+}}^{in}$ represents the in-degree of the i-th child node of node $\alpha$. $d_{\alpha}^{out}$ represents the out-degree of node $\alpha$, that is, the number of child nodes that node $\alpha$ has.

Equation Interpretation: The KQI is calculated by summing the contributions of all the parent nodes of node $\alpha$. Each parent node’s contribution is the negative logarithm of the ratio of its volume relative to node $\alpha$, weighted by the proportion of the parent node’s volume in the overall knowledge tree volume. Additionally, when calculating $V_{\alpha}^{T}$, the volume of node $\alpha$ includes not only its direct out-degree contribution but also the contribution weighted by the in-degrees of all its child nodes. This ensures the rational distribution and flow of volume within the knowledge tree. The KQI value reflects the degree of information reutilization of node $\alpha$ in the knowledge tree. A higher KQI value indicates that node $\alpha$ can reutilize information from its parent nodes more effectively, which is generally correlated with the node's importance and influence within the knowledge network.
\begin{figure}[h]
\centering 
\includegraphics[width=3.35in]{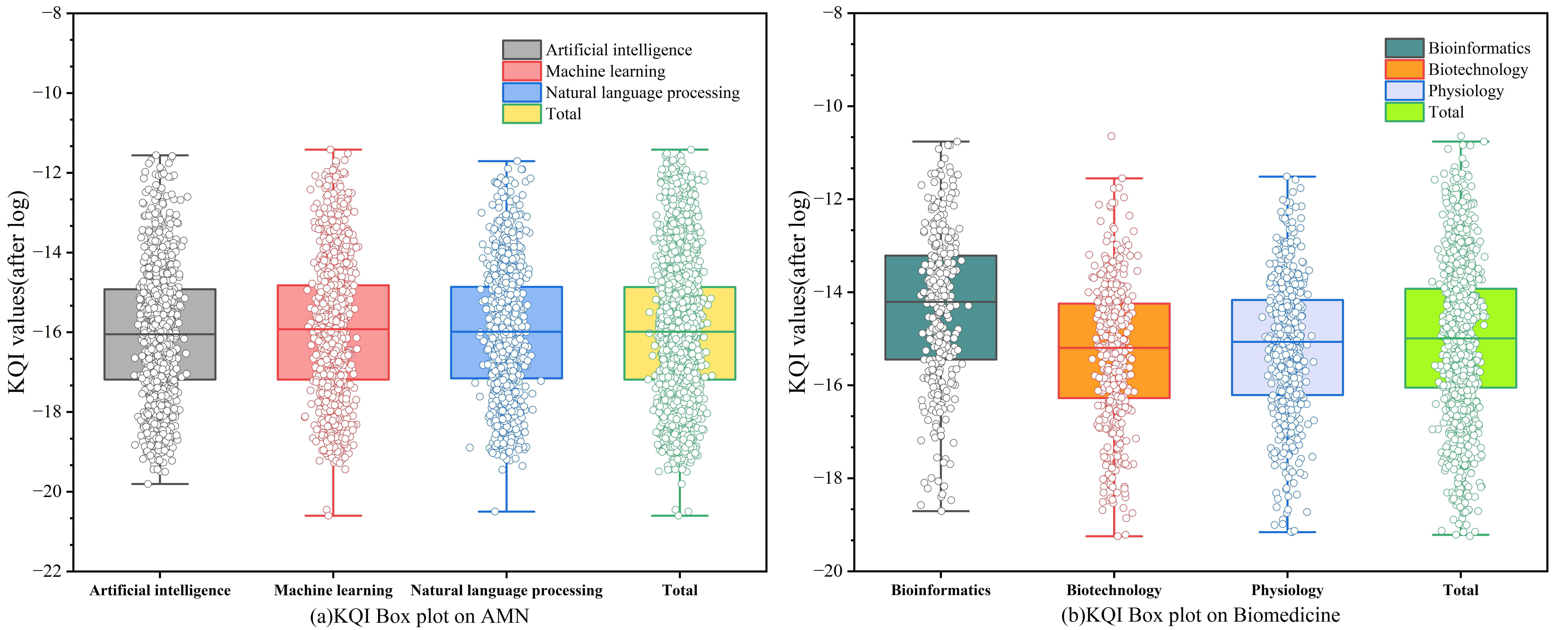}
\caption{Boxplots of KQI(after log) for two datasets}
\label{fig:kqi}
\end{figure}
\par
Based on the theoretical foundation of the Knowledge Quantification Index (KQI), the KQI value for the central node is calculated in each citation network. Figure \ref{fig:kqi} displays the distribution of KQI values, after logarithmic transformation, across two distinct domains: AMN (encompassing three subfields under computer science) and Biomedicine. The box plots provide a visual representation of knowledge quality across various subfields within these domains.

In the AMN dataset (panel a), Artificial Intelligence and Machine Learning exhibit a broader range of KQI values with medians around -14, reflecting significant variability in knowledge quality. Natural Language Processing shows a slightly more concentrated distribution, with a median also around -14, suggesting moderate consistency in knowledge quality across this subfield.

In the Biomedicine dataset (panel b), Bioinformatics stands out with the highest median KQI value, approximately -13, and a more concentrated distribution, indicating higher and more consistent knowledge quality. Biotechnology follows, with a median slightly lower than that of Bioinformatics, still suggesting robust knowledge quality. Physiology, however, presents the lowest median KQI value, around -16, with a wider spread, indicating more variable knowledge quality within this field.

The difference in KQI between each domain in the two datasets is not significant, suggesting a high degree of consistency in the overall quality of knowledge in the subdomains of the computer or biomedical fields. Consequently, the data from the two datasets are summarized and classified separately based on distinct KQI categories to evaluate the knowledge value of the articles in these fields.
\section{Result and discussion}
\begin{table*}[!t]
  \centering
  \caption{\normalfont Accuracy (ACC \%), AUC and F1 with standard deviation for node classification over 10 trials on paper citation networks. The best result is in bold, and the second best is \underline{underlined}.}
  \resizebox{\textwidth}{!}{ 
    \begin{tabular}{clcccccccccc}
    \toprule
    \multirow{2}[1]{*}{\textbf{Index}} & \multirow{2}[1]{*}{\textbf{Model}} & \multirow{2}[1]{*}{DBLPV13} & \multirow{2}[1]{*}{ALTEGRADChallenge2021} & \multirow{2}[1]{*}{ACMIS} & \multirow{2}[1]{*}{Citation Networks-V12} & \multirow{2}[1]{*}{SNAP-HEP-TH} & \multirow{2}[1]{*}{ACMHypertextECHT} & \multirow{2}[1]{*}{AI+ML+NLP(AMN)} & \multirow{2}[1]{*}{Oxytocin} & \multirow{2}[1]{*}{OGBN-ArXiv} & \multirow{2}[1]{*}{Biomedicine} \\
         &      &      &      &      &      &      &      &      &      &      &  \\
    \midrule
    \multirow{7}[0]{*}{ACC} & \textbf{Bi-LSTM+Attention} & 66.62±0.52 & 58.43±0.22 & 56.64±0.69 & 78.52±0.58 & \underline{74.52±0.41} & 84.93±0.29 & 72.68±0.62 & 70.76±0.35 & 77.30±0.22 & 74.28±0.69 \\
    & \textbf{TGNNs} & 64.33±0.23 & 57.50±0.77 & 57.30±0.19 & 67.23±0.51 & 72.16±1.23 & 60.68±0.51 & 69.01±0.34 & 65.45±0.34 & 69.81±0.63 & 71.21±0.13 \\
    & \textbf{LAD-GNN} & 68.84±0.82 & 60.75±0.46 & \underline{58.53±0.22} & 78.54±0.83 & 72.11±0.86 & 91.46±1.58 & 70.76±0.61 & 72.64±0.41 & 75.04±1.46 & \underline{80.32±0.22} \\
    & \textbf{TransformerConv\_MLP} & 64.18±1.51 & \underline{62.25±0.29} & 55.53±0.36 & \underline{81.40±0.46} & 73.52±0.47 & 73.97±0.35 & 71.35±0.27 & \underline{73.09±0.84} & 79.10±0.59 & 80.26±0.85 \\
    & \textbf{TransformerConv\_KAN} & 67.10±0.57 & 61.00±0.68 & 55.75±0.61 & 80.23±1.34 & 71.30±1.17 & 89.04±1.70 & \underline{73.68±0.95} & 68.33±0.67 & 79.62±0.83 & 77.29±0.64 \\
    & \textbf{Mamba\_MLP} & \underline{69.25±0.29} & 59.75±0.27 & 56.16±0.52 & 80.78±0.33 & 72.44±0.23 & \underline{93.15±0.98} & 71.35±0.88 & 71.45±0.52 & \underline{80.77±0.89} & 75.83±1.06 \\
    & \textbf{EMK-KEN} & \textbf{72.03±0.38} & \textbf{63.25±0.43} & \textbf{61.13±1.28} & \textbf{83.64±0.35} & \textbf{75.31±0.39} & \textbf{99.37±0.63} & \textbf{77.16±0.31} & \textbf{77.41±0.65} & \textbf{81.89±0.88} & \textbf{83.33±0.67} \\
         &      &      &      &      &      &      &      &      &      &      &  \\
    \multirow{7}[0]{*}{F1} & \textbf{Bi-LSTM+Attention} & 66.67±0.81 & 58.37±0.49 & 54.05±0.65 & 78.50±0.36 & \underline{74.51±0.75} & 84.88±0.56 & 72.72±0.63 & 71.26±0.61 & 77.31±0.24 & 74.30±0.55 \\
    & \textbf{TGNNs} & 64.33±0.38 & 56.03±0.76 & 57.18±0.48 & 67.01±0.87 & 72.08±1.14 & 60.16±0.73 & 68.34±0.17 & 64.62±0.15 & 69.79±0.51 & 71.06±0.17 \\
    & \textbf{LAD-GNN} & 68.35±0.67 & 60.40±0.63 & \underline{58.27±0.76} & 78.51±0.51 & 75.10±0.63 & 91.31±1.42 & 70.25±0.42 & 72.11±0.17 & 74.62±1.16 & \underline{80.30±0.16} \\
    & \textbf{TransformerConv\_MLP} & 64.04±1.36 & \underline{61.51±0.62} & 51.23±0.44 & \underline{81.23±0.73} & 73.41±0.64 & 71.50±0.94 & 71.18±0.45 & \underline{72.60±0.61} & 79.02±0.34 & 80.11±0.73 \\
    & \textbf{TransformerConv\_KAN} & 67.06±0.73 & 60.16±0.52 & 55.77±0.38 & 80.19±1.06 & 71.30±±0.89 & 89.15±2.26 & \underline{73.70±0.61} & 67.29±0.92 & 79.55±0.67 & 77.14±0.46 \\
    & \textbf{Mamba\_MLP} & \underline{69.34±0.40} & 59.59±0.13 & 56.12±0.31 & 80.82±0.50 & 72.21±0.34 & \underline{93.06±0.51} & 71.36±0.76 & 71.33±0.40 & \underline{80.93±0.73} & 75.72±1.15 \\
    & \textbf{EMK-KEN} & \textbf{72.11±0.65} & \textbf{63.29±0.25} & \textbf{61.26±0.95} & \textbf{83.63±0.50} & \textbf{75.34±0.42} & \textbf{99.42±0.58} & \textbf{77.19±0.28} & \textbf{77.50±0.74} & \textbf{81.90±0.72} & \textbf{83.32±0.69} \\
         &      &      &      &      &      &      &      &      &      &      &  \\
    \multirow{7}[1]{*}{AUC} & \textbf{Bi-LSTM+Attention} & 66.56±0.68 & 58.84±0.16 & 55.70±0.42 & 78.46±0.84 & \underline{74.48±0.89} & 85.16±0.14 & 72.75±0.83 & 84.56±0.89 & 94.01±0.31 & 74.34±0.66 \\
    & \textbf{TGNNs} & 64.22±0.34 & 56.76±0.52 & 57.23±0.64 & 67.48±0.34 & 71.14±1.07 & 63.83±0.35 & 67.98±0.30 & 65.53±0.12 & 89.45±0.50 & 71.54±0.24 \\
    & \textbf{LAD-GNN} & 68.40±0.52 & 60.08±0.64 & \underline{58.62±0.37} & 78.64±0.52 & 72.16±0.45 & 91.70±1.39 & 71.93±0.39 & 85.67±0.43 & 92.60±1.07 & 80.39±0.43 \\
    & \textbf{TransformerConv\_MLP} & 65.15±1.61 & \underline{61.10±0.47} & 54.75±0.13 & \underline{81.45±0.41} & 73.48±0.57 & 75.32±0.66 & 72.12±0.25 & \underline{86.00±0.55} & 93.13±0.68 & \underline{80.57±0.56} \\
    & \textbf{TransformerConv\_KAN} & 67.83±0.88 & 60.75±0.34 & 55.73±0.46 & 80.31±1.01 & 71.29±1.10 & 89.48±1.97 & \underline{74.02±0.76} & 68.25±0.77 & \underline{94.20±0.42} & 77.63±0.77 \\
    & \textbf{Mamba\_MLP} & \underline{69.71±0.52} & 59.69±0.39 & 56.08±0.47 & 80.79±0.29 & 72.37±0.58 & \underline{93.62±0.27} & 71.41±0.54 & 79.48±0.36 & 93.04±0.74 & 75.89±0.89 \\
    & \textbf{EMK-KEN} & \textbf{72.48±0.21} & \textbf{63.41±0.21} & \textbf{60.97±1.06} & \textbf{83.71±0.61} & \textbf{75.26±0.44} & \textbf{99.59±0.41} & \textbf{77.22±0.43} & \textbf{93.21±0.69} & \textbf{94.93±0.37} & \textbf{83.64±0.51} \\
    \bottomrule
    \end{tabular}%
  \label{tab:result}%
}
\end{table*}%

In this work, the EMK-KEN model was rigorously evaluated against six alternative models on a heterogeneous set of the datasets, which were categorized based on three distinct classification criteria: (1) the complexity of the citation network\cite{IdentifyingNode, snapnets, ACMHypertextECHT, ALTEGRADChallenge2021, Citation-Networks-V12, sinha2015overview, DBLP-V13, Tang:08KDD, Tang:10TKDD, Tang:11ML, Tang:12TKDE, Tang:07ICDM, DVN_27695_2014}, (2) the previous classification standards provided by the datasets\cite{OGBN-ArXiv, wang2020microsoft, mikolov2013distributed, leng_2022_6615221}, and (3) the Knowledge Quantification Index (KQI)\cite{xu2011quantifying, fanelli2019theory, ye2014approximate, coutrot2022entropy, xu2022methodology, yang2023minimum, zhang2024human, yang2024incremental, Jin2024}. The final results are summarized in Table \ref{tab:result}.

When classification was performed based on the complexity of the citation networks, the EMK-KEN model demonstrated superior performance on datasets such as DBLP-V13, Citation Networks-V12, SNAP-HEP-TH, and ACMHypertextECHT. For instance, on the Citation Networks-V12 dataset, the EMK-KEN model achieved an accuracy of 83.64\%, surpassing the mamba\_MLP model's 80.78\% and the TransformerConv\_KAN model's 80.31\%.  The complex associations within the citation networks can be effectively captured by the model, which significantly enhances its generalization ability and accuracy. Notably, on the smaller-scale like ACMHypertextECHT dataset, the performance of this model, reaching 99.37\%, is exceptionally excellent.
 
On datasets that belong to the second category, namely those with pre-existing classification standards, this model has also demonstrated impressive performance. For example, on the OGBN-ArXiv dataset, its F1 score reach to 81.90\%, which is significantly higher compared to the Bi-LSTM+Attention model's score of 77.31\%. The significant performance gap between EMK-KEN and other baseline models highlights the model's ability to efficiently handle large scaled networks, which is frequently extremely sparse with cluttered noise. Similarly, on the Oxytocin dataset, the EMK-KEN model achieved an F1 score of 77.50\% and an AUC of 93.21\%, significantly outperforming the TransformerConv\_MLP model, which obtained 72.60\% of F1 score and 86.00\% of AUC. The experimental results on Oxytocin indicate that EMK-KEN can effectively extract the complex structural features of biological networks.  Thus, the model achieves a prediction accuracy that surpasses the state-of-the-art on this type of dataset.

EMK-KEN still performs exceptionally well on the third type of classification datasets which has the Knowledge Quantification Index (KQI). For instance, on the AMN dataset, it achieved an accuracy of 77.16\%, substantially outperforming the LAD-GNN model, which 70.76\%. Meanwhile, on the Biomedicine dataset, the accuracy of EMK-KEN reached 83.33\%, while TransformerConv\_MLP reached 80.26\%. These experimental results demonstrate that the model is suitable for nodes classification based on citation networks with KQI. That means EMK-KEN's classification performance has been enhanced through the KQI of these datasets by utilizing knowledge entropy information.  
\begin{figure}[h]
\centering 
\includegraphics[width=3.40in]{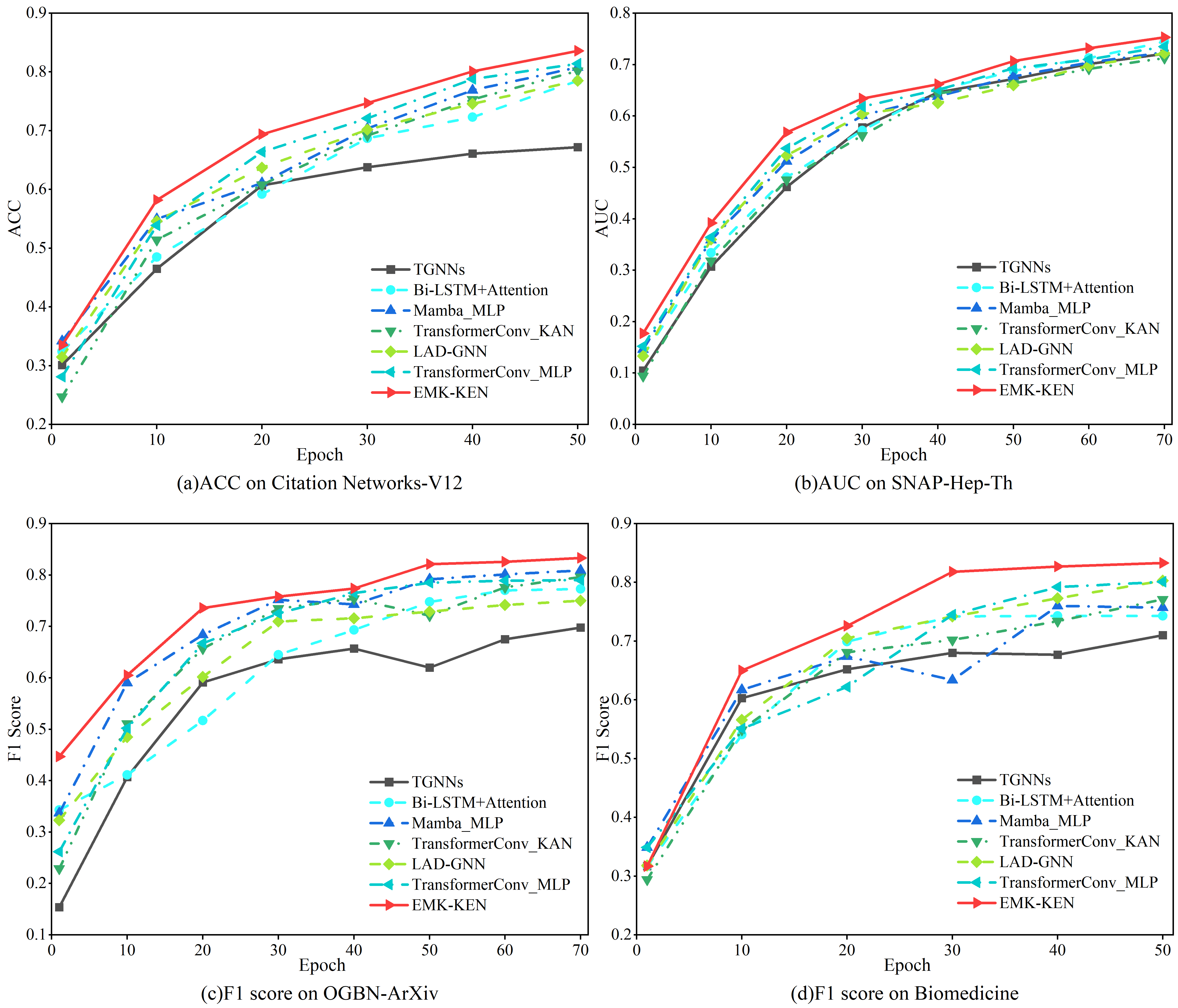}
\caption{Performance comparison of different models on multiple datasets}
\label{fig:result1}
\end{figure}

\begin{figure}[h]
\centering 
\includegraphics[width=3.40in]{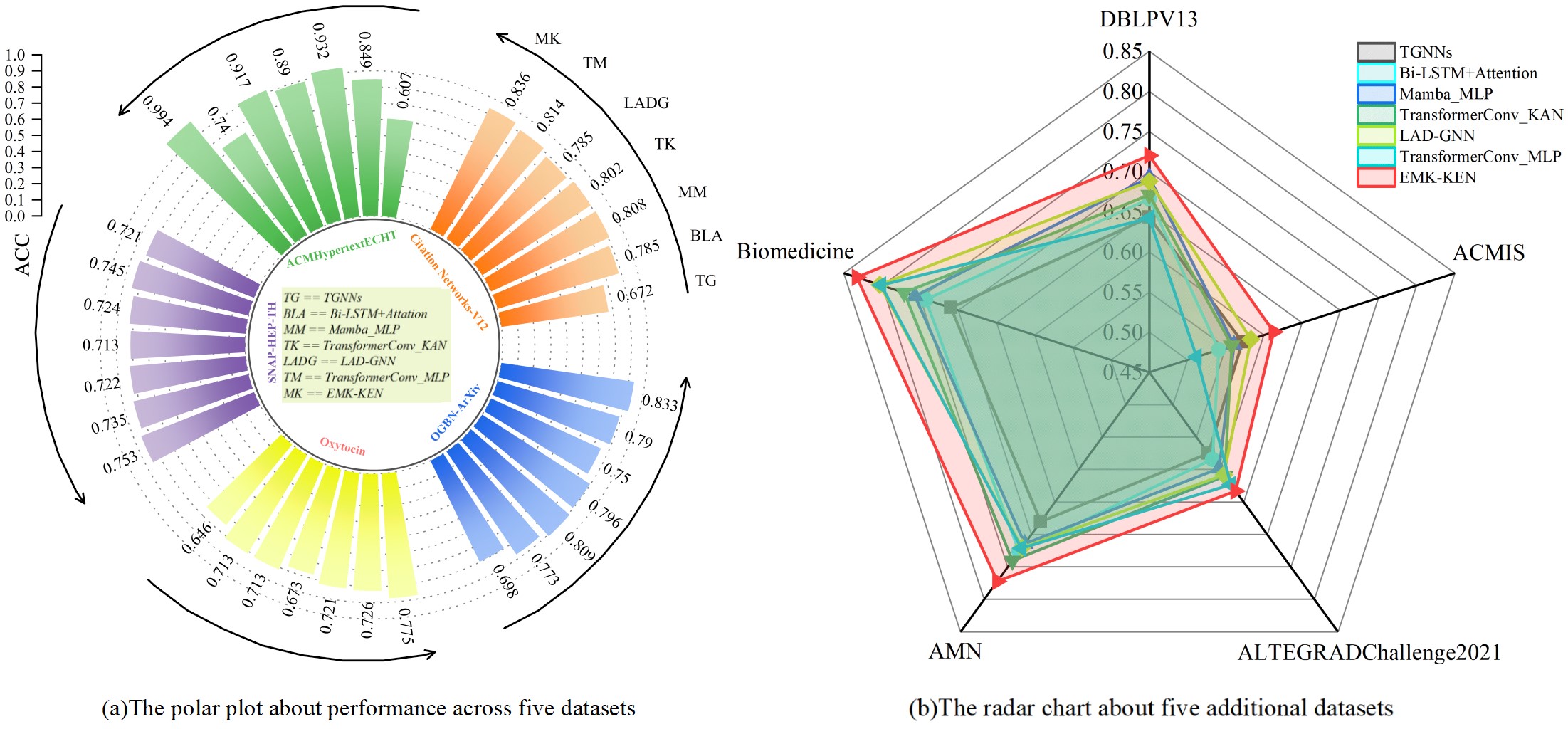}
\caption{The final performance of different models on the datasets}
\label{fig:result2}
\end{figure}
Additionally, to further evaluate the efficiency and stability of the EMK-KEN during the training process, line charts were plotted to display the changes in the F1 scores, accuracy (ACC), and AUC of these models across different datasets as the number of training epochs increases. As shown in Figure \ref{fig:result1}, it is evident that, regardless of the classification method or evaluation metric used, the EMK-KEN model is able to rapidly capture the structural information present in the citation network.

For instance, on the Citation Networks-V12 dataset (Figure \ref{fig:result1}(a)), the EMK-KEN model maintains a leading position in accuracy throughout the entire training process, with other models, such as Bi-LSTM+Attention and TGNNs, notably lagging behind. On the SNAP-HEP-TH dataset (Figure \ref{fig:result1}(b)), the EMK-KEN model exhibits the highest AUC growth rate, particularly in the early training stages, where its AUC increases more rapidly than that of other models and stabilizes in the subsequent training phases. This indicates its strong ability to learn deep features in complex networks. On the OGBN-ArXiv dataset (Figure \ref{fig:result1}(c)), the EMK-KEN model's F1 score rises rapidly, ultimately reaching 0.82, surpassing both TGNNs and LAD-GNN, highlighting its precision under existing classification standards. On the Biomedicine dataset (Figure \ref{fig:result1}(d)), the use of KQI allows the model to more effectively quantify the knowledge entropy in the graph structure during the feature extraction and classification process. When KQI is used as the classification standard, the F1 score of the EMK-KEN model rises to approximately 0.84, consistently outperforming all six other models at each training stage. This demonstrates its superior adaptability and higher classification accuracy. The model effectively captures and utilizes the complex knowledge information in the graph structure, achieving more accurate classification results.

Furthermore, Figure \ref{fig:result2} presents a detailed visualization of the performance of seven models across ten datasets, utilizing two subgraphs. The polar plot on the left illustrates the performance across five datasets—Citation Networks-V12, SNAP-HEP-TH, OGBN-ArXiv, ACMHypertextECHT, and Oxytocin. Each axis of the plot represents the performance metrics of the models on these datasets. The colored fan-shaped regions represent various models, such as TGNNs, Bi-LSTM+Attention, Mamba\_MLP, among others, visually illustrating the performance differences of each model across the datasets. On the right, the radar chart compares the performance of the models across five additional datasets: DBLPV13, ACMIS, ALTEGRADChallenge2021, Biomedicine, and AMN. Areas of different colors represent each model, enabling a clear comparison of their performance on these datasets. For example, the EMK-KEN model outperformed other models on the DBLPV13 dataset, whereas, on different datasets, different models exhibit varying strengths. This detailed visualization emphasizes the adaptability of the EMK-KEN model and highlights its consistent dominance across a wide range of citation networks and datasets.

The experimental results shown in Figures \ref{fig:result1} and \ref{fig:result2} illustrate that the EMK-KEN model possesses superior generalization ability and robustness. This indicates that the model is efficient and accurate in extracting complex structural features from citation networks. \\

\begin{fontsize}{12pt}{15pt} 
\textbf{Hyperparameter Analysis}
\end{fontsize}
\begin{figure}[h]
\centering 
\includegraphics[width=3.40in]{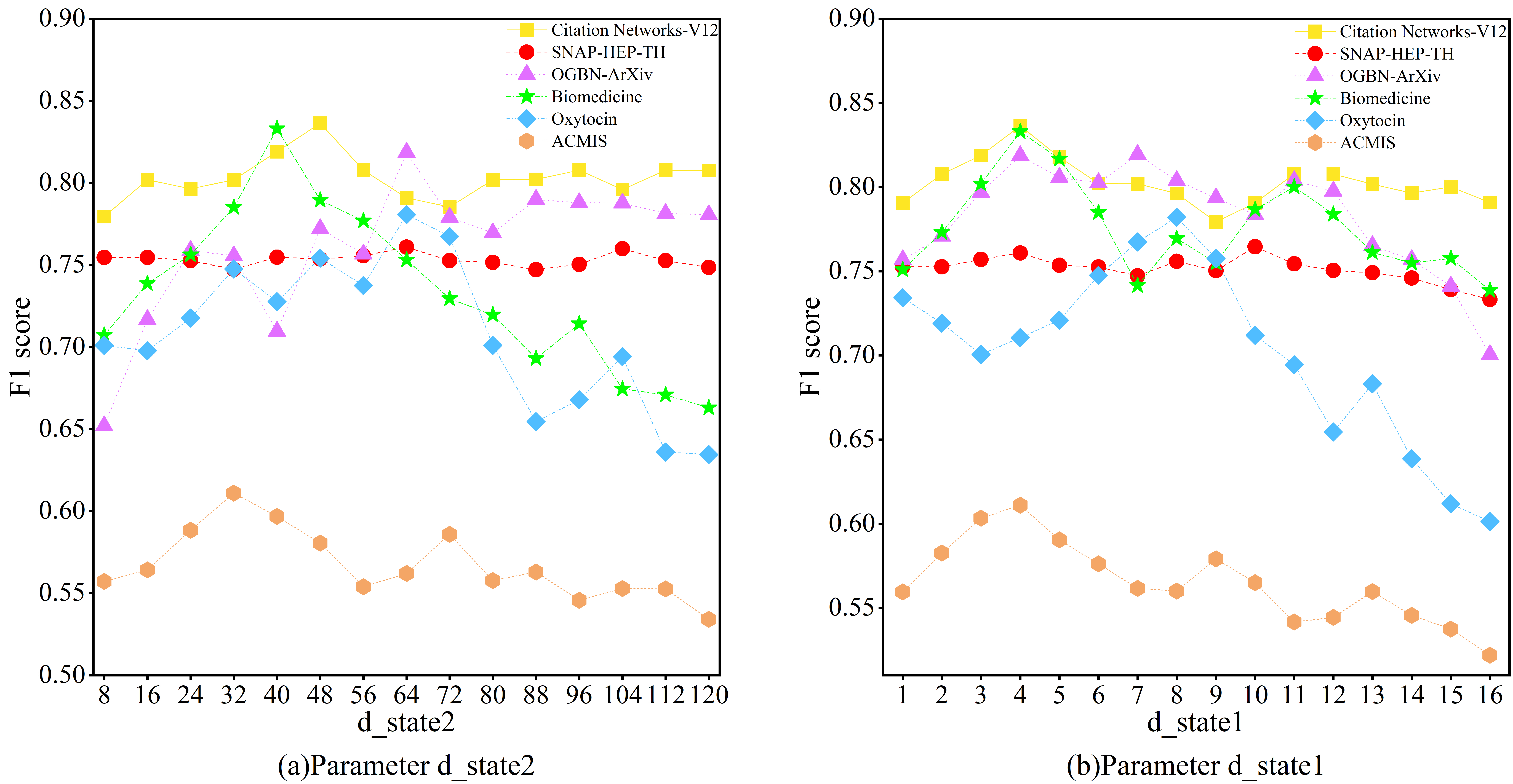}
\caption{Hyperparameter sensitiveness of d\_state on the six datasets.}
\label{fig:d-state}
\end{figure}

\begin{table*}[!t]
  \centering
  \caption{\normalfont Ablation experiments on six datasets}
      \resizebox{\textwidth}{!}{ 
    \begin{tabular}{clcccccc}
    \toprule
    \multirow{2}[2]{*}{\textbf{Index}} & \multirow{2}[2]{*}{\textbf{Model}} & \multirow{2}[2]{*}{\textbf{DBLP-V13}} & \multirow{2}[2]{*}{\textbf{SNAP-HEP-TH}} & \multirow{2}[2]{*}{\textbf{Citation Networks-V12}} & \multirow{2}[2]{*}{\textbf{OGBN-ArXiv}} & \multirow{2}[2]{*}{\textbf{Biomedicine}} & \multirow{2}[2]{*}{\textbf{Oxytocin}} \\
    &      &      &      &      &      &      &  \\
    \midrule
    \multirow{7}[1]{*}{ACC} & \textbf{-MetaFP} & ---  & ---  & 78.34±0.49 & ---  & 78.86±0.58 & 70.10±0.76 \\
    & \textbf{-Causal convolution} & 66.89±0.42 & 62.65±0.09 & 70.24±0.53 & 65.20±0.95 & 62.10±0.36 & 45.85±1.28 \\
    & \textbf{-SSMS} & 65.82±0.75 & 64.91±0.26 & 75.26±0.81 & 67.04±0.66 & 77.34±0.19 & 71.76±0.21 \\
    & \textbf{-Mamba Block} & 63.70±1.27 & 68.52±0.42 & 74.15±0.38 & 51.01±0.83 & 75.83±0.24 & 69.44±0.45 \\
    & \textbf{-KAN} & 64.82±0.51 & 70.54±0.13 & 74.60±0.35 & 74.74±0.52 & 71.19±0.42 & 64.28±0.61 \\
    & \textbf{-Dropout layer after KAN} & 67.51±0.90 & 73.11±0.28 & 79.91±1.04 & 78.30±0.77 & 75.85±0.79 & 68.18±0.10 \\
    & \textbf{Whole model} & \textbf{70.93±0.85} & \textbf{76.20±0.57} & \textbf{82.43±0.36} & \textbf{84.12±0.46} & \textbf{81.51±0.33} & \textbf{73.68±0.27} \\
    &      &      &      &      &      &      &  \\
    \multirow{7}[0]{*}{F1} & \textbf{-MetaFP} & ---  & ---  & 78.31±0.62 & ---  & 78.61±0.53 & 70.77±0.64 \\
    & \textbf{-Causal convolution} & 66.96±0.43 & 62.55±0.18 & 70.23±0.25 & 65.17±1.03 & 61.83±0.17 & 44.20±1.36 \\
    & \textbf{-SSMS} & 65.81±0.70 & 64.63±0.39 & 75.20±1.14 & 66.91±0.72 & 77.26±0.14 & 70.87±0.87 \\
    & \textbf{-Mamba Block} & 63.68±1.14 & 68.37±0.16 & 74.11±0.88 & 51.62±0.94 & 75.77±0.36 & 69.14±0.49 \\
    & \textbf{-KAN} & 64.85±0.63 & 70.62±0.25 & 74.37±0.37 & 74.83±0.41 & 71.02±0.38 & 64.57±0.52 \\
    & \textbf{-Dropout layer after KAN} & 67.61±0.65 & 72.88±0.33 & 79.94±1.10 & 78.64±0.95 & 75.28±0.52 & 68.45±0.37 \\
    & \textbf{Whole model} & \textbf{70.76±0.98} & \textbf{76.14±0.24} & \textbf{82.12±0.45} & \textbf{84.24±0.23} & \textbf{81.24±0.35} & \textbf{73.38±0.16} \\
    &      &      &      &      &      &      &  \\
    \multirow{7}[1]{*}{AUC} & \textbf{-MetaFP} & ---  & ---  & 78.43±0.16 & ---  & 79.08±0.42 & 83.69±0.31 \\
    & \textbf{-Causal convolution} & 67.03±0.55 & 63.54±0.12 & 70.29±0.39 & 76.83±1.27 & 62.42±0.41 & 58.13±1.27 \\
    & \textbf{-SSMS} & 66.67±0.54 & 65.18±0.45 & 75.33±0.94 & 78.14±0.52 & 77.48±0.25 & 84.43±0.42 \\
    & \textbf{-Mamba Block} & 64.43±1.35 & 68.80±0.23 & 74.02±0.72 & 67.55±1.01 & 75.90±0.47 & 82.22±0.36 \\
    & \textbf{-KAN} & 65.21±0.46 & 70.89±0.37 & 74.23±0.21 & 88.35±0.39 & 71.54±0.35 & 76.78±0.79 \\
    & \textbf{-Dropout layer after KAN} & 67.93±0.78 & 73.12±0.34 & 80.04±0.93 & 89.21±1.06 & 76.21±0.76 & 80.11±0.19 \\
    & \textbf{Whole model} & \textbf{71.54±1.02} & \textbf{76.08±0.32} & \textbf{82.71±0.24} & \textbf{93.46±0.15} & \textbf{81.73±0.24} & \textbf{86.62±0.14} \\
    \bottomrule
    \end{tabular}%
  \label{tab:ablation}%
  }
\end{table*}%

\par

The critical parameter $d\_state$, which influences the state matrices $\boldsymbol{A}$, $\boldsymbol{B}$, $\boldsymbol{C}$, and the state vector in equations \ref{eq:6}, \ref{eq:7}, and \ref{eq:8}, is further discussed. Since metadata and text embeddings are processed separately through Mamba block, two key parameters are involved: $d\_state1$ (used for processing metadata) and $d\_state2$ (used for processing text embeddings). While keeping other hyperparameters constant, $d\_state1$ is initially set to a specific value, and the value of $d\_estate2$ is gradually adjusted from 8 to 120. Subsequently, the classification performance of the EMK-KEN was tested on different datasets. As depicted in Figure \ref{fig:d-state}(a), it can be observed that different values of $d\_state2$ have a significant impact on the model's performance, indicating that the model is relatively sensitive to $d\_state2$. Additionally, for different datasets, distinct optimal values of $d\_state2$ exist to ensure that the model's performance remains consistently high. Various values of $d\_state2$ are carefully experimented and then the appropriate values for each dataset are determined. Next, based on the selected $d\_state2$, the value of $d\_state1$ is adjusted and incremented  from 1 to 16 and the classification performance of the model is tested on different datasets. As shown in Figure \ref{fig:d-state}(b), it can be observed that the model is also relatively sensitive to $d\_state1$. After testing, the suitable values of $d\_state1$ are determined for each dataset.
\\

\begin{fontsize}{12pt}{15pt} 
\textbf{Ablation Study}
\end{fontsize}
\par
Table \ref{tab:ablation} presents the results of a comprehensive ablation study on the EMK-KEN model, systematically assessing the contribution of each architectural component. The study involved sequentially removing the MetaFP layer, causal convolution layer, state space models (SSMS), Mamba block, KAN layer, and the Dropout layer following the KAN layer.
\begin{itemize}
\item The MetaFP layer, which is responsible for metadata processing, had a significant impact on performance when removed. On the Citation Networks-V12, Biomedicine, and Oxytocin datasets, accuracies decreased by 2.65\%-4.09\%, underscoring its importance in feature extraction.
\item The causal convolution layer was similarly crucial. Its removal caused a sharp drop in performance, particularly on the Oxytocin dataset, where the ACC score decreased by 27.83\% and the F1 score dropped by 29.18\%. This underscores its role in modeling causal relationships and information dissemination.
\item The SSMS, a key component of the Mamba architecture, was essential for maintaining high performance on dynamic datasets. On the DBLP-V13 and SNAP-HEP-TH datasets, performance decreased by 5.11\% and 11.29\%, respectively, without it, emphasizing its role in handling such data.
\item The Mamba block, which is essential for handling long sequences, exhibited a significant reduction in performance when removed. The OGBN-ArXiv dataset results, which showed a 33.11\% decrease in accuracy, underscore its role in managing long-range dependencies.
\item The KAN layer, which enhances the model's representational capacity, was similarly important. Its removal led to performance decreases across multiple datasets, with an average accuracy drop of 8.12\%.
\item Notably, the removal of the Dropout layer following the KAN layer also impacted performance. On the Citation Networks-V12 and OGBN-ArXiv datasets, accuracies decreased by 3.33\% and 5.82\%, respectively, suggesting its role in preventing overfitting in specific scenarios.
\end{itemize}
\par
These ablation studies demonstrate that the design of the components and their collaborative cooperation is both necessary and effective.

\section{Conclusion}
 EMK-KEN, a novel knowledge evaluation approach, has been proposed with lower computational complexity, stronger generalization capabilities, and higher precision compared to the state-of-the-art algorithm. The Mamba layer, a component of the network, adeptly processes attributes associated with nodes in citation networks. It excels in handling long-sequence text embeddings, facilitates linear processing, and demonstrates proficiency in recognizing intricate patterns. The learnable activation function of the KAN layer improves the ability of the model to capture structural information. Experiments conducted on ten benchmark datasets, using three evaluation indicators and comparing performance with six other models, demonstrate superior results. This research lays the foundation for future literature analysis and knowledge mining. Future efforts may focus on enhancing the model's scalability and adaptability for broader knowledge evaluation and analysis tasks.

\bibliographystyle{ieeetr}
\bibliography{bare_conf_compsoc}

\begin{thebibliography}{10}

\bibitem{HOU2019100197}
J.~Hou, H.~Pan, T.~Guo, I.~Lee, X.~Kong, and F.~Xia, ``Prediction methods and applications in the science of science: A survey,'' {\em Computer Science Review}, vol.~34, p.~100197, Nov. 2019.

\bibitem{AnIndexToQuantify}
J.~E. Hirsch, ``An index to quantify an individual's scientific research output,'' {\em Proceedings of the National Academy of Sciences of the United States of America}, vol.~102, pp.~16569--16572, Nov. 2005.

\bibitem{garfield1972citation}
E.~Garfield, ``Citation analysis as a tool in journal evaluation: Journals can be ranked by frequency and impact of citations for science policy studies.,'' {\em Science}, vol.~178, pp.~471--479, Nov. 1972.

\bibitem{yan2011citation}
R.~Yan, J.~Tang, X.~Liu, D.~Shan, and X.~Li, ``Citation count prediction: learning to estimate future citations for literature,'' in {\em Proceedings of the 20th ACM international conference on Information and knowledge management}, pp.~1247--1252, 2011.

\bibitem{co-citationanalysis}
C.~M. Trujillo and T.~M. Long, ``Document co-citation analysis to enhance transdisciplinary research,'' {\em Science Advances}, vol.~4, no.~1, p.~e1701130, 2018.

\bibitem{anil2020effect}
A.~Anil and S.~R. Singh, ``Effect of class imbalance in heterogeneous network embedding: An empirical study,'' {\em Journal of Informetrics}, vol.~14, p.~101009, May 2020.

\bibitem{qin2017dual}
Y.~Qin, D.~Song, H.~Chen, W.~Cheng, G.~Jiang, and G.~W. Cottrell, ``A dual-stage attention-based recurrent neural network for time series prediction,'' in {\em Proceedings of the Twenty-Sixth International Joint Conference on Artificial Intelligence, {IJCAI-17}}, pp.~2627--2633, Aug. 2017.

\bibitem{zhang2020predicting}
F.~Zhang and S.~Wu, ``Predicting future influence of papers, researchers, and venues in a dynamic academic network,'' {\em Journal of Informetrics}, vol.~14, p.~101035, May 2020.

\bibitem{li2022predicting}
X.~Li, X.~Tang, and Q.~Cheng, ``Predicting the clinical citation count of biomedical papers using multilayer perceptron neural network,'' {\em Journal of Informetrics}, vol.~16, p.~101333, Nov. 2022.

\bibitem{3637871}
J.~Gao, X.~Zhao, M.~Li, M.~Zhao, R.~Wu, R.~Guo, Y.~Liu, and D.~Yin, ``Smlp4rec: An efficient all-mlp architecture for sequential recommendations,'' {\em ACM Trans. Inf. Syst.}, vol.~42, Jan. 2024.

\bibitem{Wang2025}
X.~Wang, X.~Ao, F.~Zhang, Z.~Zhang, and Q.~He, ``Knowledge error detection via textual and structural joint learning,'' {\em Big Data Mining and Analytics}, vol.~8, pp.~233--240, Feb. 2025.

\bibitem{10026520}
X.~Wu, J.~Duan, Y.~Pan, and M.~Li, ``Medical knowledge graph: Data sources, construction, reasoning, and applications,'' {\em Big Data Mining and Analytics}, vol.~6, no.~2, pp.~201--217, 2023.

\bibitem{he2023h2cgl}
G.~He, Z.~Xue, Z.~Jiang, Y.~Kang, S.~Zhao, and W.~Lu, ``H2cgl: Modeling dynamics of citation network for impact prediction,'' {\em Information Processing \& Management}, vol.~60, p.~103512, Nov. 2023.

\bibitem{yu2021incorporating}
Y.~Yu, G.~Wang, H.~Ren, and Y.~Cai, ``Incorporating bidirection-interactive information and semantic features for relational facts extraction (student abstract),'' in {\em Proceedings of the AAAI Conference on Artificial Intelligence}, vol.~35, pp.~15947--15948, May 2021.

\bibitem{zou2023se}
D.~Zou, H.~Peng, X.~Huang, R.~Yang, J.~Li, J.~Wu, C.~Liu, and P.~S. Yu, ``Se-gsl: A general and effective graph structure learning framework through structural entropy optimization,'' in {\em Proceedings of the ACM Web Conference 2023}, pp.~499--510, Apr. 2023.

\bibitem{duan2024structural}
L.~Duan, X.~Chen, W.~Liu, D.~Liu, K.~Yue, and A.~Li, ``Structural entropy based graph structure learning for node classification,'' in {\em Proceedings of the AAAI Conference on Artificial Intelligence}, vol.~38, pp.~8372--8379, Mar. 2024.

\bibitem{hong2024label}
X.~Hong, W.~Li, C.~Wang, M.~Lin, and S.~Lu, ``Label attentive distillation for gnn-based graph classification,'' in {\em Proceedings of the AAAI Conference on Artificial Intelligence}, vol.~38, pp.~8499--8507, Mar. 2024.

\bibitem{corso2024graph}
G.~Corso, H.~Stark, S.~Jegelka, T.~Jaakkola, and R.~Barzilay, ``Graph neural networks,'' {\em Nature Reviews Methods Primers}, vol.~4, p.~17, Mar. 2024.

\bibitem{zhou2020graph}
J.~Zhou, G.~Cui, S.~Hu, Z.~Zhang, C.~Yang, Z.~Liu, L.~Wang, C.~Li, and M.~Sun, ``Graph neural networks: A review of methods and applications,'' {\em AI open}, vol.~1, pp.~57--81, Jan. 2021.

\bibitem{gamarnik2023barriers}
D.~Gamarnik, ``Barriers for the performance of graph neural networks (gnn) in discrete random structures,'' {\em Proceedings of the National Academy of Sciences of the United States of America}, vol.~120, p.~e2314092120, Oct. 2023.

\bibitem{uddin2024}
A.~M. ud~din and S.~Qureshi, ``Limits of depth: Over-smoothing and over-squashing in gnns,'' {\em Big Data Mining and Analytics}, vol.~7, pp.~205--216, Mar. 2024.

\bibitem{SchiffKGDGK24}
Y.~Schiff, C.-H. Kao, A.~Gokaslan, T.~Dao, A.~Gu, and V.~Kuleshov, ``Caduceus: Bi-directional equivariant long-range dna sequence modeling,'' in {\em ICML}, 2024.

\bibitem{gu2023mamba}
A.~Gu and T.~Dao, ``Mamba: Linear-time sequence modeling with selective state spaces,'' {\em CoRR}, vol.~abs/2312.00752, Dec. 2023.

\bibitem{behrouz2024graph}
A.~Behrouz and F.~Hashemi, ``Graph mamba: Towards learning on graphs with state space models,'' in {\em Proceedings of the 30th ACM SIGKDD Conference on Knowledge Discovery and Data Mining}, pp.~119--130, Aug. 2024.

\bibitem{dao2024transformers}
T.~Dao and A.~Gu, ``Transformers are {SSM}s: Generalized models and efficient algorithms through structured state space duality,'' in {\em Proceedings of the 41st International Conference on Machine Learning} (R.~Salakhutdinov, Z.~Kolter, K.~Heller, A.~Weller, N.~Oliver, J.~Scarlett, and F.~Berkenkamp, eds.), vol.~235 of {\em Proceedings of Machine Learning Research}, pp.~10041--10071, PMLR, 21--27 Jul 2024.

\bibitem{SHUKLA2024117290}
K.~Shukla, J.~D. Toscano, Z.~Wang, Z.~Zou, and G.~E. Karniadakis, ``A comprehensive and fair comparison between mlp and kan representations for differential equations and operator networks,'' {\em Computer Methods in Applied Mechanics and Engineering}, vol.~431, p.~117290, 2024.

\bibitem{liu2024kan}
Z.~Liu, Y.~Wang, S.~Vaidya, F.~Ruehle, J.~Halverson, M.~Solja{\v{c}}i{\'c}, T.~Y. Hou, and M.~Tegmark, ``Kan: Kolmogorov-arnold networks,'' {\em arXiv preprint arXiv:2404.19756}, June 2024.

\bibitem{liu2024kan2}
Z.~Liu, P.~Ma, Y.~Wang, W.~Matusik, and M.~Tegmark, ``Kan 2.0: Kolmogorov-arnold networks meet science,'' {\em arXiv preprint arXiv:2408.10205}, Aug. 2024.

\bibitem{bresson2024kagnns}
R.~Bresson, G.~Nikolentzos, G.~Panagopoulos, M.~Chatzianastasis, J.~Pang, and M.~Vazirgiannis, ``Kagnns: Kolmogorov-arnold networks meet graph learning,'' {\em arXiv preprint arXiv:2406.18380}, July 2024.

\bibitem{kiamari2024gkan}
M.~Kiamari, M.~Kiamari, and B.~Krishnamachari, ``Gkan: Graph kolmogorov-arnold networks,'' {\em arXiv preprint arXiv:2406.06470}, June 2024.

\bibitem{yang2024kolmogorov}
X.~Yang and X.~Wang, ``Kolmogorov-arnold transformer,'' {\em arXiv preprint arXiv:2409.10594}, Sept. 2024.

\bibitem{beltagy2019scibert}
I.~Beltagy, K.~Lo, and A.~Cohan, ``Scibert: A pretrained language model for scientific text,'' in {\em Proceedings of the 2019 Conference on Empirical Methods in Natural Language Processing and the 9th International Joint Conference on Natural Language Processing (EMNLP-IJCNLP)} (K.~Inui, J.~Jiang, V.~Ng, and X.~Wan, eds.), (Hong Kong, China), pp.~3615--3620, Association for Computational Linguistics, Nov. 2019.

\bibitem{ma2020charbert}
W.~Ma, Y.~Cui, C.~Si, T.~Liu, S.~Wang, and G.~Hu, ``Charbert: Character-aware pre-trained language model,'' in {\em Proceedings of the 28th International Conference on Computational Linguistics} (D.~Scott, N.~Bel, and C.~Zong, eds.), (Barcelona, Spain (Online)), pp.~39--50, International Committee on Computational Linguistics, Dec. 2020.

\bibitem{gupta2022matscibert}
T.~Gupta, M.~Zaki, N.~A. Krishnan, and Mausam, ``Matscibert: A materials domain language model for text mining and information extraction,'' {\em npj Computational Materials}, vol.~8, p.~102, May 2022.

\bibitem{Zhang2021}
J.~Zhang and Q.~Xu, ``Attention-aware heterogeneous graph neural network,'' {\em Big Data Mining and Analytics}, vol.~4, pp.~233--241, Dec. 2021.

\bibitem{snapnets}
J.~Leskovec and A.~Krevl, ``{SNAP Datasets}: {Stanford} large network dataset collection.'' \url{http://snap.stanford.edu/data}, June 2014.

\bibitem{ACMHypertextECHT}
Anderson, ``In-conference citation data acm hypertext \& echt (version 2).'' \url{http://dx.doi.org/10.5258/SOTON/D1870v2}, July 2021.

\bibitem{ALTEGRADChallenge2021}
K.~Assobo, ``Altegrad challenge 2021.'' \url{https://www.kaggle.com/datasets/kevinassobo/altegrad-challenge-2021/data}, June 2021.

\bibitem{Citation-Networks-V12}
J.~Tang, ``Citation-networks-v12.'' \url{https://www.kaggle.com/datasets/mathurinache/citation-network-dataset}, Apr. 2020.

\bibitem{sinha2015overview}
A.~Sinha, Z.~Shen, Y.~Song, H.~Ma, D.~Eide, B.-j.~P. Hsu, and K.~Wang, ``An overview of microsoft academic service (mas) and applications,'' in {\em Proceedings of the 24th international conference on world wide web}, pp.~243--246, ACM, May 2015.

\bibitem{DBLP-V13}
J.~Tang, ``Dblp-citation-network v13.'' \url{https://www.aminer.cn/citation}, May 2021.

\bibitem{Tang:08KDD}
J.~Tang, J.~Zhang, L.~Yao, J.~Li, L.~Zhang, and Z.~Su, ``Arnetminer: Extraction and mining of academic social networks,'' in {\em KDD'08}, pp.~990--998, Aug. 2008.

\bibitem{Tang:10TKDD}
J.~Tang, L.~Yao, D.~Zhang, and J.~Zhang, ``A combination approach to web user profiling,'' {\em ACM TKDD}, vol.~5, pp.~1--44, Dec. 2010.

\bibitem{Tang:11ML}
J.~Tang, J.~Zhang, R.~Jin, Z.~Yang, K.~Cai, L.~Zhang, and Z.~Su, ``Topic level expertise search over heterogeneous networks,'' {\em Machine Learning Journal}, vol.~82, pp.~211--237, Sept. 2010.

\bibitem{Tang:12TKDE}
J.~Tang, A.~C. Fong, B.~Wang, and J.~Zhang, ``A unified probabilistic framework for name disambiguation in digital library,'' {\em IEEE Transactions on Knowledge and Data Engineering}, vol.~24, pp.~975--987, June 2012.

\bibitem{Tang:07ICDM}
J.~Tang, D.~Zhang, and L.~Yao, ``Social network extraction of academic researchers,'' in {\em ICDM'07}, pp.~292--301, Oct. 2007.

\bibitem{DVN_27695_2014}
P.~Luo, ``{ACM IS abstract and citation network1},'' 2014.

\bibitem{OGBN-ArXiv}
S.~University, ``Ogbn-arxiv (processed for pyg).'' \url{https://ogb.stanford.edu/docs/nodeprop/\#ogbn-arxiv}, May 2020.

\bibitem{wang2020microsoft}
K.~Wang, Z.~Shen, C.~Huang, C.-H. Wu, Y.~Dong, and A.~Kanakia, ``Microsoft academic graph: When experts are not enough,'' {\em Quantitative Science Studies}, vol.~1, pp.~396--413, Feb. 2020.

\bibitem{mikolov2013distributed}
T.~Mikolov, I.~Sutskever, K.~Chen, G.~S. Corrado, and J.~Dean, ``Distributed representations of words and phrases and their compositionality,'' in {\em Advances in Neural Information Processing Systems} (C.~Burges, L.~Bottou, M.~Welling, Z.~Ghahramani, and K.~Weinberger, eds.), vol.~26, Curran Associates, Inc., 2013.

\bibitem{leng_2022_6615221}
R.~I. Leng, ``Citation network data sets for 'oxytocin – a social peptide? deconstructing the evidence','' June 2022.

\bibitem{IdentifyingNode}
S.~Huang, T.~Lv, X.~Zhang, Y.~Yang, W.~Zheng, and C.~Wen, ``Identifying node role in social network based on multiple indicators,'' {\em PLOS ONE}, vol.~9, pp.~1--16, Aug. 2014.

\bibitem{xu2011quantifying}
Y.~Xu and A.~Bernard, ``Quantifying the value of knowledge within the context of product development,'' {\em Knowledge-Based Systems}, vol.~24, pp.~166--175, Feb. 2011.

\bibitem{fanelli2019theory}
D.~Fanelli, ``A theory and methodology to quantify knowledge,'' {\em \textit{R. Soc. Open Sci.}}, vol.~6, p.~181055, Apr. 2019.

\bibitem{ye2014approximate}
C.~Ye, R.~C. Wilson, C.~H. Comin, L.~d.~F. Costa, and E.~R. Hancock, ``Approximate von neumann entropy for directed graphs,'' {\em Physical Review E}, vol.~89, p.~052804, May 2014.

\bibitem{coutrot2022entropy}
A.~Coutrot, E.~Manley, S.~Goodroe, C.~Gahnstrom, G.~Filomena, D.~Yesiltepe, R.~C. Dalton, J.~M. Wiener, C.~H{\"o}lscher, M.~Hornberger, {\em et~al.}, ``Entropy of city street networks linked to future spatial navigation ability,'' {\em Nature}, vol.~604, pp.~104--110, Mar. 2022.

\bibitem{xu2022methodology}
H.~Xu, R.~Luo, J.~Winnink, C.~Wang, and E.~Elahi, ``A methodology for identifying breakthrough topics using structural entropy,'' {\em Information Processing \& Management}, vol.~59, p.~102862, Mar. 2022.

\bibitem{yang2023minimum}
Z.~Yang, G.~Zhang, J.~Wu, J.~Yang, Q.~Z. Sheng, H.~Peng, A.~Li, S.~Xue, and J.~Su, ``Minimum entropy principle guided graph neural networks,'' in {\em Proceedings of the sixteenth ACM international conference on web search and data mining}, pp.~114--122, Feb. 2023.

\bibitem{Jin2024}
X.~Jin, F.~Zhu, Y.~Shen, G.~Jeon, and D.~Camacho, ``Data-driven dynamic graph convolution transformer network model for eeg emotion recognition under iomt environment,'' {\em Big Data Mining and Analytics}, 2024.

\bibitem{zhang2024human}
C.~Zhang, B.~Jia, Y.~Zhu, and S.-C. Zhu, ``Human-level few-shot concept induction through minimax entropy learning,'' {\em Science Advances}, vol.~10, p.~eadg2488, Apr. 2024.

\bibitem{yang2024incremental}
R.~Yang, H.~Peng, C.~Liu, and A.~Li, ``Incremental measurement of structural entropy for dynamic graphs,'' {\em Artificial Intelligence}, vol.~334, p.~104175, Sept. 2024.

\end{thebibliography}

\end{document}